\documentclass[english]{article}
\usepackage[T1]{fontenc}
\usepackage[latin9]{inputenc}
\usepackage{geometry}
\usepackage{amsmath}
\usepackage{graphicx}
\usepackage{hyperref}

\makeatletter

\usepackage{cite}

\newcommand\rmd{\mathrm{d}}
\date{}

\let\oldbibliography\thebibliography
\renewcommand{\thebibliography}[1]{%
  \small%
  \oldbibliography{#1}%
  \setlength{\itemsep}{0pt}%
}

\makeatother

\usepackage{babel}
\begin{document}
\title{Supportive interactions in the noisy voter model}
\author{A. Kononovicius\thanks{email: \protect\href{mailto:aleksejus.kononovicius@tfai.vu.lt}{aleksejus.kononovicius@tfai.vu.lt};
website: \protect\url{http://kononovicius.lt}}}
\date{Institute of Theoretical Physics and Astronomy, Vilnius University}
\maketitle
\begin{abstract}
Latane social impact theory predicts recruitment and supportive interactions
being responsible for opinion formation. So far only recruitment interactions
were considered in the voter models. Here we consider a noisy voter model with
supportive interactions, which make voters less likely to change their
opinions. This is similar to the voter models with freezing, but instead
of interacting with their past selves voters get the support from
their peers. We examine two different ways in which the support could
be implemented: support deterring imitation as well as independence,
support deterring imitation only. Both assumptions introduce strong
drift into the model, which almost always overcomes the diffusion
caused by the imitative behavior. The first assumption introduces
strong attraction to a full consensus state, unless the support becomes
too strong. The latter assumption promotes partial consensus, with
surviving minority group.
\end{abstract}

\section{Introduction}

It is widely accepted that humans tend to become more alike as they
intermingle. Yet somehow human societies remains heterogeneous: holding
a variety of different beliefs, customs and opinions \cite{Axelrod1997JConfRes,Flache2017JASSS}.
Sometimes heterogeneity of the commonly held beliefs appears to be
contradicting even scientific consensus on relatively simple matters
\cite{Binder2009CommRes,Galam2010PhysA,McCright2011SocQ,Schmidt2018Vac,Hardy2019SciComm}.
Phenomenology and mechanisms of individual and collective opinion
adoption have been investigated both from the socio--psychological
point of view \cite{Willis1965,Latane1981,Akerlof2009Princeton,Lilleker2014}
as well as by developing social models similar to the ones found in
physics \cite{Castellano2009RevModPhys,Stauffer2013JStatPhys,Abergel2017Springer,Baronchelli2018RSOS,Jedrzejewski2019CRP,Redner2019CRP,Galesic2019PhysA}.
Most models considered in sociophysics involve binary opinions \cite{Castellano2009RevModPhys,Stauffer2013JStatPhys,Abergel2017Springer,Baronchelli2018RSOS,Jedrzejewski2019CRP,Redner2019CRP,Galesic2019PhysA},
though some multi--state generalizations of the binary models are
also considered (see \cite{Kononovicius2017Complexity,Bancerowski2019EPJB,Vazquez2019PRE}
for some recent examples). Prevalence of binary opinion models can
be likely attributed to the heritage of the Ising model. Furthermore
assuming binary opinions is often sufficient to reproduce major social
phenomena. In contrast, computational social science seems to prefer
models with continuous opinions, such as bounded confidence models
\cite{Flache2017JASSS}.

Voter model is one of the most studied agent--based models in sociophysics
\cite{Castellano2009RevModPhys,Stauffer2013JStatPhys,Baronchelli2018RSOS,Abergel2017Springer,Jedrzejewski2019CRP,Redner2019CRP,Galesic2019PhysA}.
While the original voter model \cite{Clifford1973} assumed competition
between species, this simple imitation model has found its applications
in opinion dynamics \cite{Liggett1999}. Original voter model has
numerous generalizations, which explore the impact of variety of social
interaction mechanisms. Some of the more well known generalizations
include independent (or noisy) behavior \cite{Granovsky1995}, inflexibility
\cite{Mobilia2007JStatMech,Khalil2018PRE}, anti--conformity \cite{Tanabe2013Chaos,Krueger2017Entropy,Khalil2019PhysA},
variety of interaction (network) topologies \cite{Alfarano2009Dyncon,Kononovicius2014EPJB},
non--linear interactions \cite{Castellano2009PRE,Peralta2018Chaos}
and even memory \cite{Stark2008PRL,Stark2008ACS,Wang2014SciRep,Artime2018PRE}.
Voter model has also found applications outside the field of opinion
dynamics, e.g., it was used as a base for the models of the financial
markets \cite{Alfarano2005CompEco,Alfarano2008Dyncon,Kononovicius2012PhysA,Gontis2014PlosOne,Lux2018JEDC}.

Here we consider the effect of the supportive interactions on the
noisy voter model. The noisy voter model is particularly interesting
model as it involves two main social responses (as per the diamond
model \cite{Willis1965}) to the peer pressure: independence (referred
to as variability in the diamond model) and conformity. The diamond
model also predicts two additional social responses: anti--conformity
and disregard (referred to as independence in the diamond model).
Disregarding the peer pressure evidently does not change the system
state, thus this mechanism would have no observable effect. Anti--conformity,
on the other hand, would change the system state, but it can be easily
shown that it is already included in the noisy voter model. Introducing
anti--conformity simply readjusts model parameter values unless some
additional assumptions are made, such as considering conformists and
anti--conformists being two different agent types as in \cite{Tanabe2013Chaos,Khalil2019PhysA},
or assuming network topology with at least two cliques as in \cite{Krueger2017Entropy}.
It is worth to note that these social responses have profound impact
in other generalizations of the voter model, such as $q$-voter model
\cite{Nail2016APPA}.

Another well--known psychological theory is Latane social impact
theory \cite{Latane1981}. While this theory is mostly concerned with
the magnitude of impact social forces have on the individuals, this
theory unlike the diamond model \cite{Willis1965} acknowledges the
possibility of positive interactions between individuals holding similar
beliefs. Namely it suggests that interactions with like minded individuals
can trigger not only anti--conformity as the diamond model predicts,
but also grant higher resistance to persuasion by the individuals
with different beliefs. Notably Latane social impact theory has a
dedicated binary opinion agent--based model \cite{Nowak1990PsychRev},
which was recently generalized to account for the non--binary opinions
\cite{Bancerowski2019EPJB}. This model, while being conceptually
similar, is noticeably more complicated (e.g., it includes long--range
spatial interactions) than the voter model. Thus unlike the voter
model it does not allow for analytical interpretation. Therefore here
we explore how supportive interactions, predicted by the Latane social
impact theory \cite{Latane1981}, but absent from earlier generalizations
of the voter model, influence the dynamics of the noisy voter model.

Supportive interactions, imagined as resistance to persuasion, is
similar to the opinion freezing previously considered in \cite{Stark2008PRL,Stark2008ACS,Wang2014SciRep}
as in both cases transition rates decrease under the considered effect.
Our approach differs in that the resistance is conditioned not on
the time the opinion is held, but on the support provided by other
agents in the current moment in time. In other words, one could say
that opinion freezing is effectively agents obtaining support from
their past selves, while in our approach agents obtain support in
real time from their like--minded peers. Hence our approach is purely
Markovian, while the model considered in \cite{Stark2008PRL,Stark2008ACS,Wang2014SciRep}
is non--Markovian. Having to deal with Markovian model is convenient
for us as we can derive some of the results analytically to verify
the results obtained from numerical simulations. In the absence of
noise, similarity could be also drawn to the majority vote model \cite{Oliveira1992JStatPhys,Liggett1999,Vilela2018SciRep}
as the transition will occur only if there is more agents encouraging
transition than ones providing support the current beliefs. Though
in our approach the transition rate is not constant as in the majority
vote models, but is proportional to the difference between the numbers
of agents holding each opinion.

This paper is organized as follows. In Section~\ref{sec:vm} we provide
a brief description of the noisy voter model and the two different
agent interaction scenarios we consider in this paper. In Section~\ref{sec:model1}
we introduce the first model with supportive interactions. The first
model assumes that support deters both independence and imitation.
The second model assumes that support deters only imitation and is
discussed in Section~\ref{sec:model2}. A short summary of our findings
and conclusions are provided in Section~\ref{sec:Conclusions}.

\section{Extensive and non--extensive noisy voter models\label{sec:vm}}

Original formulation of the model, which is now known as the voter
model, included only the copying mechanism \cite{Clifford1973,Liggett1999}.
In the original model a randomly selected agent simply copies the
state of his randomly selected neighbor. The original model was concerned
with competition between species instead of competition between social
behaviors, but in social scenarios the copying mechanism can be seen
to represent social imitation or conforming to the perceived norm.
Yet the conformist response is not the only possible response to the
peer pressure: one can also disregard the opinions of the other agents,
act contrary to the opinions of the other agents, keep or change the
opinion independently \cite{Willis1965}. While these social responses
have major impact on the dynamics of the $q$-voter models \cite{Nail2016APPA},
in the original voter model only conformity and independence seem
to be important. Model involving these two mechanisms is known as
the noisy voter model \cite{Granovsky1995}. In general case we can
write down the transition rates of the noisy voter model as follows:
\begin{align}
\lambda & \left(X\rightarrow X+1\right)=\lambda^{+}=\left(N-X\right)\left\{ \sigma_{1}+\frac{h}{N^{\alpha}}X\right\} ,\nonumber \\
\lambda & \left(X\rightarrow X-1\right)=\lambda^{-}=X\left\{ \sigma_{0}+\frac{h}{N^{\alpha}}\left(N-X\right)\right\} .\label{eq:vm-rates}
\end{align}
In the above $N$ is the total number of agents in the system, $X$
is the number of agents in the state $1$, $\sigma_{i}$ are independent
transition rates to the state $i$ and $h$ is the peer imitation
rate. Later we will be considering infinite $N$ limit, in that case
it is reasonable to introduce scaled system state variable $x=\frac{X}{N}$,
which corresponds to the fraction of agents in the state $1$. These
rates assume that the agents are able to interact with every other
agent.

Parameter $\alpha$ encodes the weight of independence: with $\alpha=0$
the agents see themselves to be as important as any one of their peers,
on the other hand if $\alpha=1$, then the agents see themselves to
be as important as all of their peers combined. In \cite{Alfarano2005CompEco,Alfarano2008Dyncon,Alfarano2009Dyncon}
$\alpha=1$ case is referred to as the local interactions, while $\alpha=0$
case is referred to as the global interactions. In \cite{Kononovicius2014EPJB}
it was shown that by continuously changing $\alpha$ we observe continuous
transition between non--extensive ($\alpha=0$) and extensive ($\alpha=1$)
statistical description of the system. For finitely large $N$ this
distinction seems to be redundant as one can simply bypass it by appropriately
rescaling $h$ value. Yet in the infinite $N$ limit there is a profound
difference: if $\alpha=1$, then the discrete model is approximated
by ordinary differential equation (abbr. ODE), on the other hand if
$\alpha=0$, then the discrete model is approximated by stochastic
differential equation (abbr. SDE).

Approximation of the discrete model can be obtained by noting that
the rates describe one--step transitions, which can be seen to describe
generation (birth) and recombination (death) of the particles (agents).
Thus we can use the birth--death process formalism \cite{VanKampen2007NorthHolland}
to obtain the following general formula:
\begin{equation}
\rmd x=\frac{\lambda^{+}-\lambda^{-}}{N}\rmd t+\sqrt{\frac{\lambda^{+}+\lambda^{-}}{N^{2}}}\rmd W.\label{eq:gen-sde}
\end{equation}
In the above $W$ is the standard one dimensional Brownian motion
(also known as the Wiener process).

By putting the transition rates, Eq.~(\ref{eq:vm-rates}), into Eq.~(\ref{eq:gen-sde})
for the finitely large $N$ we obtain:
\begin{equation}
\rmd x=\left\{ \sigma_{1}\left(1-x\right)-\sigma_{0}x\right\} \rmd t+\sqrt{\frac{2h}{N^{\alpha}}x\left(1-x\right)+\frac{1}{N}\left\{ \sigma_{1}\left(1-x\right)+\sigma_{0}x\right\} }\rmd W.
\end{equation}
In the infinite $N$ limit we would drop any terms of the order $N^{-\alpha}$
with $\alpha>0$ as those terms quickly approach $0$. Thus the diffusion
function would disappear for $\alpha>0$ and the process would be
described by an ODE. It should be evident that the ODE has a stable
fixed point at:
\begin{equation}
x=\frac{\sigma_{1}}{\sigma_{0}+\sigma_{1}}.
\end{equation}
For $\alpha=1$ the diffusion function is non--zero and the process
is described by an SDE. The steady state of the SDE is not a fixed
point, but a stationary distribution. In this particular case the
stationary distribution is the Beta distribution with the following
parameters:
\begin{equation}
x\sim\mathcal{B}e\left(\frac{\sigma_{1}}{h},\text{\ensuremath{\frac{\sigma_{0}}{h}}}\right).
\end{equation}
As the behavior of the noisy voter model depends on the ratio between
$\sigma_{i}$ and $h$, let us keep the value of $h$ fixed, $h=1$.

Let us introduce anti--conformity into the noisy voter model. In
the simplest case, when all agents exhibit both conformist and anti--conformist
responses, the transition rates have the following form:
\begin{align}
\lambda^{+}= & \left(N-X\right)\left\{ \sigma_{1}+\frac{h}{N^{\alpha}}X+\frac{g}{N^{\alpha}}\left(N-X\right)\right\} ,\nonumber \\
\lambda^{-}= & X\left\{ \sigma_{0}+\frac{h}{N^{\alpha}}\left(N-X\right)+\frac{g}{N^{\alpha}}X\right\} .\label{eq:vm-rates-ac}
\end{align}
In the above $g$ is the rate of anti-conformity. By rearranging the
terms in the transition rates we see that anti--conformity does not
introduce a qualitative change into the model (as long as $h>g>0$):
\begin{align}
\lambda^{+}= & \left(N-X\right)\left\{ \sigma_{1}+gN^{1-\alpha}+\frac{h-g}{N^{\alpha}}X\right\} =\left(N-X\right)\left\{ \sigma_{1}^{\prime}+\frac{h^{\prime}}{N^{\alpha}}X\right\} ,\nonumber \\
\lambda^{-}= & X\left\{ \sigma_{0}+gN^{1-\alpha}+\frac{h-g}{N^{\alpha}}\left(N-X\right)\right\} =X\left\{ \sigma_{0}^{\prime}+\frac{h^{\prime}}{N^{\alpha}}\left(N-X\right)\right\} .\label{eq:vm-rates-ac-1}
\end{align}
This and other cases, such as having separate types of agents for
conformists and anti--conformists, of the noisy voter model with
anti--conformity were studied in detail in \cite{Khalil2019PhysA}.
It was reported that independent transitions are no longer required
to observe seemingly independent transitions. This is true as for
$\sigma_{i}=0$ we still have $\sigma_{i}^{\prime}>0$ due to anti--conformity.
Also in case with two separate agent types (conformists and anti--conformists)
small number of anti--conformists are able to change the stationary
distribution of the model with finite $N$. 

\section{Support deterring independence and imitation\label{sec:model1}}

Let agents in the same state (allies) provide support to each other
and thus discourage switch away from current state. First let us assume
that this support discourages both independence and imitation. We
encode this assumption by considering the following transition rates:
\begin{align}
\lambda^{+} & \left(X\right)=\left(N-X\right)\left[\sigma_{1}+\frac{h}{N^{\alpha}}X-\frac{q}{N^{\beta}}\left(N-X\right)\right]_{+},\nonumber \\
\lambda^{-} & \left(X\right)=X\left[\sigma_{0}+\frac{h}{N^{\alpha}}\left(N-X\right)-\frac{q}{N^{\beta}}X\right]_{+}.\label{eq:pa-rates}
\end{align}
In the above the special brackets are equivalent to the following
$\max\left(\ldots\right)$ function: 
\begin{equation}
\left[z\right]_{+}=\max\left(z,0\right).
\end{equation}

\begin{figure}
\begin{centering}
\includegraphics[width=0.7\textwidth]{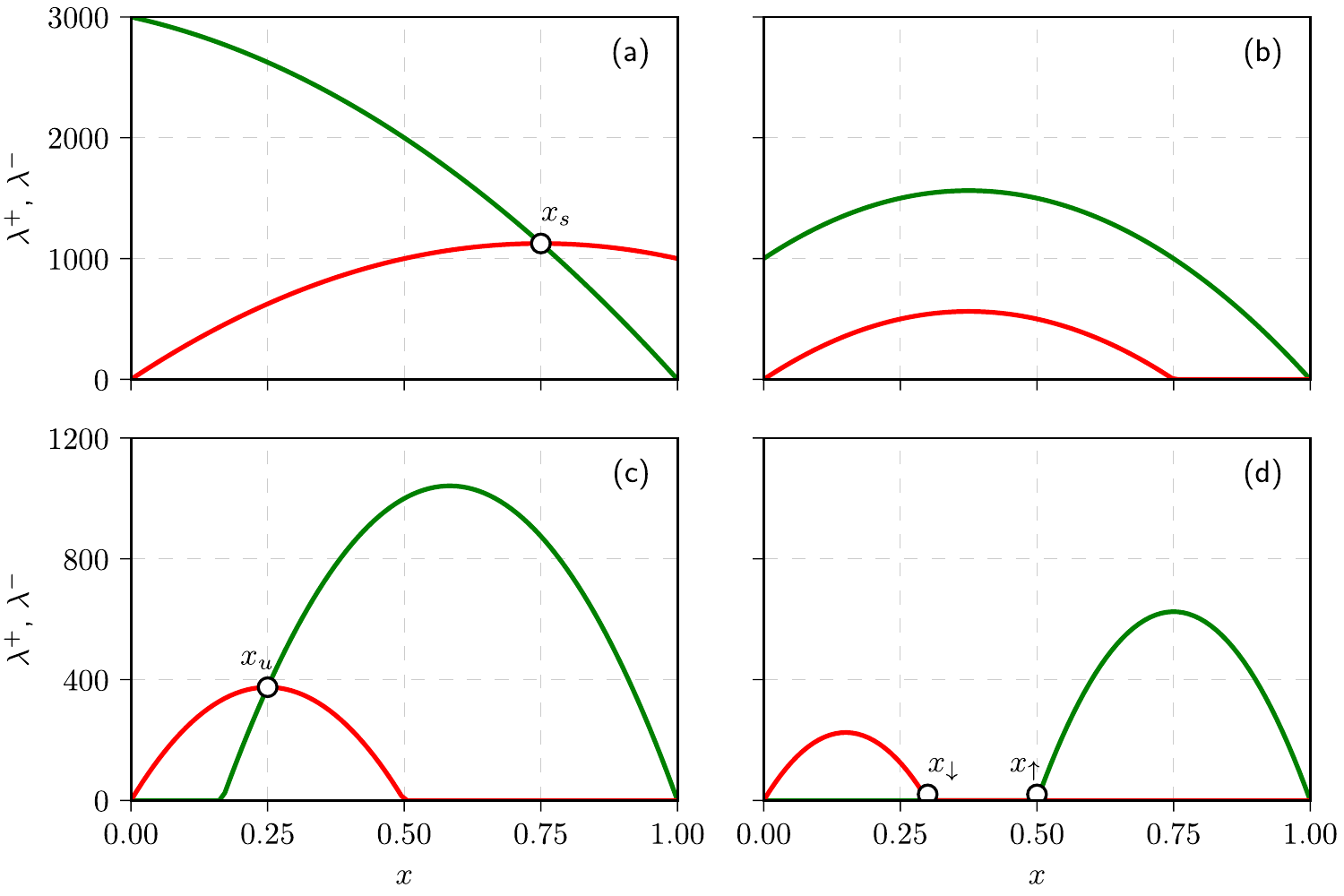}
\par\end{centering}
\caption{Four different regimes of the model observed by comparing the birth
(green curve) and the death (red curve) transition rates, Eq.~(\ref{eq:pa-rates}).
Open circles mark fixed points of the model. Model parameters: $N=10^{3}$,
$h=1$, $\sigma_{0}=2$, $\sigma_{1}=4$, $\alpha=\beta=1$ (all cases),
$q=1$ (a), $3$ (b), $5$ (c) and $9$ (d).}

\label{fig:pa-regimes}
\end{figure}

Close examination of the transition rates, Eq.~(\ref{eq:pa-rates}),
suggests that the model has four different regimes (examples of which
are shown in Fig.~\ref{fig:pa-regimes}):
\begin{itemize}
\item In the first regime (Fig.~\ref{fig:pa-regimes} (a)) both of the
special brackets in the transition are positive for all $X\in\left[0,N\right]$.
\item In the second regime (Fig.~\ref{fig:pa-regimes} (b)) one of the
special brackets is zero for some $X\in\left[0,N\right]$, while the
other remains positive for all $X\in\left[0,N\right]$.
\item In the third regime (Fig.~\ref{fig:pa-regimes} (c)) both of the
special brackets are zero for some $X\in\left[0,N\right]$, but the
set of $X$, for which $\lambda^{+}\left(X\right)>0$ and $\lambda^{-}\left(X\right)>0$,
is not empty (i.e., the transition rates overlap).
\item In the fourth regime (Fig.~\ref{fig:pa-regimes} (d)) both of the
special brackets are zero for some $X\in\left[0,N\right]$ and the
set of $X$, for which $\lambda^{+}\left(X\right)>0$ and $\lambda^{-}\left(X\right)>0$,
is empty (i.e., the transition rates do not overlap).
\end{itemize}

The transition points between these regimes, or the critical points,
are found whenever both rates go to zero at the same $X$. Thus the
critical points can be determined by solving the following system
of equations 
\begin{equation}
\begin{cases}
\lambda^{+}\left(xN\right)=0\\
\lambda^{-}\left(xN\right)=0
\end{cases},\label{eq:pa-qeqs}
\end{equation}
in respect to $x$ and $q$. While solving Eq.~(\ref{eq:pa-qeqs})
we treat the special brackets $\left[z\right]_{+}$ as if they were
ordinary ones. With this caveat we obtain three solutions of Eq.~(\ref{eq:pa-qeqs}):
\begin{align}
x_{1}=0 & ,\ensuremath{\quad q_{1}=N^{\beta-1}\sigma_{1},}\\
x_{2}=1 & ,\quad q_{2}=N^{\beta-1}\sigma_{0},\\
x_{3}=\frac{Nh+N^{\alpha}\sigma_{0}}{2Nh+N^{\alpha}\left(\sigma_{0}+\sigma_{1}\right)} & ,\quad q_{3}=N^{\beta-\alpha-1}\left(Nh+N^{\alpha}\left(\sigma_{0}+\sigma_{1}\right)\right).\label{eq:pa-q3}
\end{align}
Note that $q_{3}$ will be always larger than $q_{1}$ and $q_{2}$.
Therefore the the third critical point, $q_{c,3}$, corresponds to
$q_{3}$. The first two critical points correspond to $q_{1}$ and
$q_{2}$. Which of them is smaller depends on the selected values
of $\sigma_{0}$ and $\sigma_{1}$. For the asymmetric case, $\sigma_{0}\neq\sigma_{1}$,
we have:
\begin{align}
q_{c,1} & =N^{\beta-1}\min\left(\sigma_{0},\sigma_{1}\right),\quad q_{c,2}=N^{\beta-1}\max\left(\sigma_{0},\sigma_{1}\right).
\end{align}
In the symmetric case, $\sigma_{0}=\sigma_{1}$, we have $q_{1}=q_{2}$
and thus the second regime would not be observed.

\begin{figure}
\begin{centering}
\includegraphics[width=0.9\textwidth]{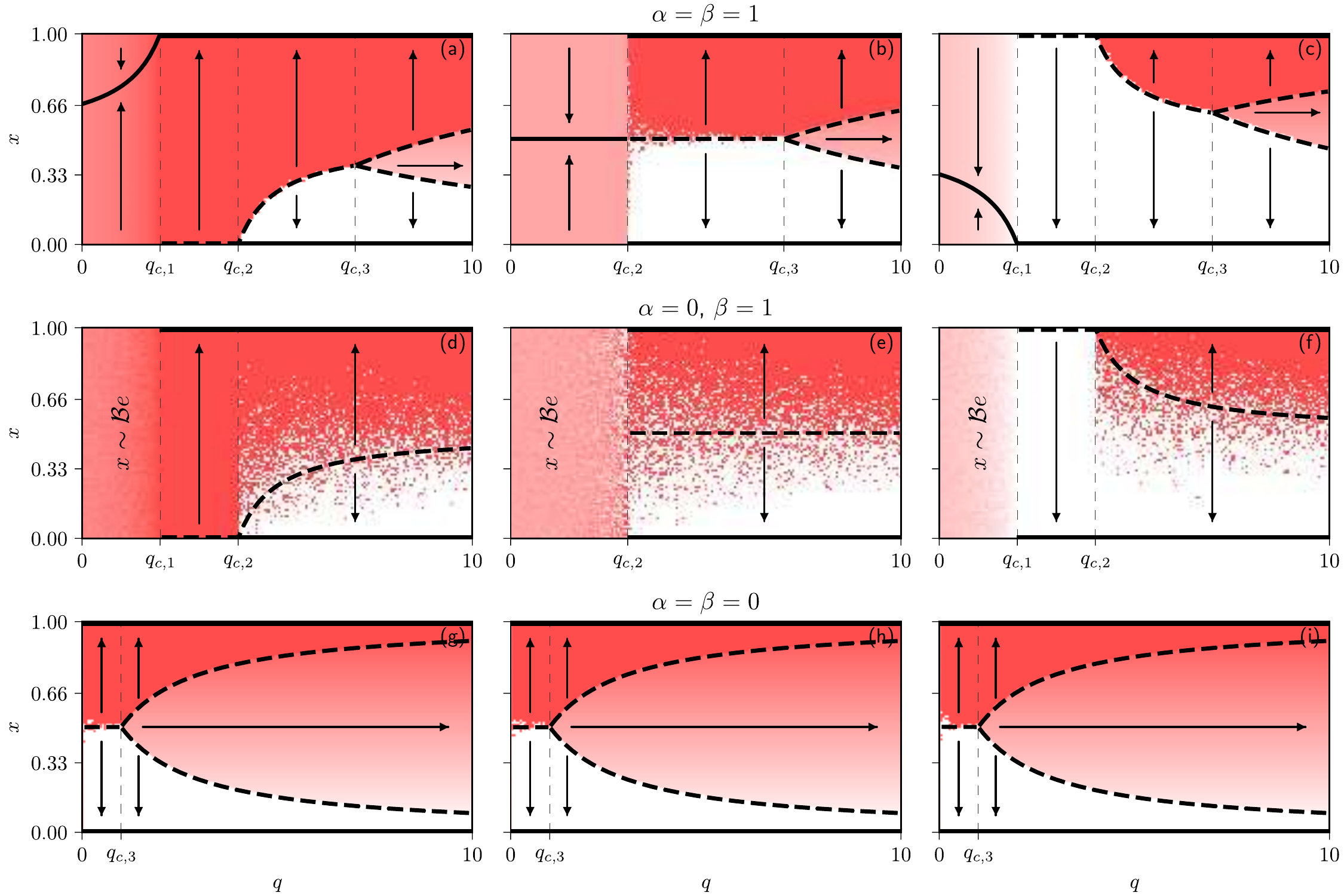}
\par\end{centering}
\caption{Bifurcation diagrams of the model driven by Eq.~(\ref{eq:pa-rates})
in respect to the intensity of support, $q$. Black curves correspond
to stable fixed points (solid curves) and unstable fixed points (dashed
curves). Arrows indicate directions in which the model moves. Areas
are shaded according to the value of $x$ reached after sufficiently
long time (white corresponds to $x\left(t\right)=0$, red corresponds
to $x\left(t\right)=1$, while shades correspond to some intermediate
values). Model parameters used to produce this figure: $N=10^{3}$
and $h=1$ (all cases), $\sigma_{0}=2$ and $\sigma_{1}=4$ ((a),
(d) and (g)), $\sigma_{0}=\sigma_{1}=3$ ((b), (e) and (h)), $\sigma_{0}=4$
and $\sigma_{1}=2$ ((c), (f) and (i)), $\alpha=\beta=1$ ((a)--(c)),
$\alpha=0$ and $\beta=1$ ((d)--(f)), $\alpha=\beta=0$ ((g)--(i)).}
\label{fig:pa-phases}
\end{figure}

In Fig.~\ref{fig:pa-phases} we have shown the dependence of the
fixed points on intensity of support, $q$, in the model driven by
Eq.~(\ref{eq:pa-rates}). We have considered extensive ($\alpha=\beta=1$),
non--extensive ($\alpha=\beta=0$) and hybrid ($\alpha=0$ and $\beta=1$)
interaction cases. The extensive case and the non--extensive case
assume that all agents interact in the same manner (either extensive
or non--extensive), while the hybrid case assumes that the agents
in different state (opposition) have more influence. Assuming the
opposite, that allies have more influence (i.e., setting $\alpha=1$
and $\beta=0$), might be a more reasonable assumption in the opinion
dynamics context, but it leads to a trivial model as both rates quickly
go to zero with larger $N$.

Let us continue with a discussion of Fig.~\ref{fig:pa-phases} and
obtain the expressions for the fixed points. For $q<q_{c,1}$ we observe
either a stable fixed point or a stationary distribution. Stable fixed
point $x_{s}$, which is observed only for $\alpha=\beta=1$, has
to satisfy:
\begin{equation}
\lambda^{+}\left(x_{s}N\right)=\lambda^{-}\left(x_{s}N\right).
\end{equation}
This fixed point was highlighted in Fig.~\ref{fig:pa-regimes} (a).
We once again treat the special brackets $\left[z\right]_{+}$ as
if they were ordinary and obtain:
\begin{equation}
x_{s}=\frac{q-\sigma_{1}}{2q-\left(\sigma_{0}+\sigma_{1}\right)}.\label{eq:pa-xs}
\end{equation}
For $\alpha=0$ and finite $N$ we observe the Beta--binomial distribution,
in the infinite $N$ limit the stationary distribution converges to
the Beta distribution:
\begin{equation}
x\sim\mathcal{B}e\left(\frac{\sigma_{1}-qN^{1-\beta}}{h+qN^{-\beta}},\frac{\sigma_{0}-qN^{1-\beta}}{h+qN^{-\beta}}\right).
\end{equation}
Earlier works \cite{Alfarano2009Dyncon,Kononovicius2014EPJB} have
already observed the emergence of the stationary Beta distribution
in the non--extensive voter model ($\alpha=0$ and $q=0$) and its
convergence to the Dirac delta function as $\alpha\rightarrow1$ and
$N\rightarrow\infty$. The result we have obtained here complements
the previous knowledge and shows that the stationary distribution
can be also observed for $q>0$. Furthermore the result has an interesting
implication from opinion dynamics perspective. Namely, a society of
independent (informed or educated) individuals who provide strong
support for their peers can be statistically indistinguishable from
a society thriving on imitation (with individuals being uninformed
or badly educated). 

In all asymmetric cases the second regime is observed. In the second
regime, for $q_{c,1}<q<q_{c,2}$, we observe fixed points at $x=0$
and $x=1$: one of them being stable, the other being unstable. Which
is which depends on the parameters $\sigma_{0}$ and $\sigma_{1}$.
If $\sigma_{0}<\sigma_{1}$, then $x=1$ is the stable fixed point
and $x=0$ is the unstable fixed point. In the opposite case, $\sigma_{0}>\sigma_{1}$,
$x=0$ is the stable fixed point and $x=1$ is the unstable fixed
point.

Special transitional regime is observed for $\alpha=\beta=1$ with
$\sigma_{0}=\sigma_{1}=q$. In this case the model does not experience
drift, only diffusion is present if $N$ is finite. Therefore for
finite $N$ the model randomly diffuses until it sticks to either
$x=0$ or $x=1$. In the infinite $N$ limit diffusion function also
goes to zero and all points become stable.

In all cases for $q_{c,2}<q<q_{c,3}$ we observe two ordered phases
are separated by an unstable fixed point. This unstable fixed point
$x_{u}$ has to satisfy:
\begin{equation}
\lambda^{+}\left(x_{u}N\right)=\lambda^{-}\left(x_{u}N\right).
\end{equation}
This fixed point was highlighted in Fig.~\ref{fig:pa-regimes} (c).
We once again treat the special brackets $\left[z\right]_{+}$ as
if they were ordinary and obtain:
\begin{equation}
x_{u}=\frac{Nq-N^{\beta}\sigma_{1}}{2Nq-N^{\beta}\left(\sigma_{0}+\sigma_{1}\right)}.
\end{equation}
Note that this expression is similar to the one obtained for $x_{s}$,
only the considered $q$ value range is different. Indeed if we set
$\beta=1$, we would obtain exactly the same expression as in Eq.~(\ref{eq:pa-xs}).

For $\alpha=0$ the separation between the phases is not deterministic,
but probabilistic instead. This is because in this case the contribution
of diffusion is non--negligible even if we take infinite $N$ limit.
Therefore a process starting below $x_{u}$ has a non--zero probability
to be absorbed at $x=1$. The opposite, process starting above $x_{u}$
and being absorbed at $x=0$ with non--zero probability, is also
true in this case. Therefore the numerical results shown in Fig.~\ref{fig:pa-phases}
appear to be somewhat scattered. Especially around the $x_{u}$ curve.

In all cases for $q_{c,3}<q$ we observe a set of stable fixed points
in the middle (marked by the right arrow in Fig.~\ref{fig:pa-phases}).
The boundaries of this region are obtained by solving
\begin{equation}
\lambda^{+}\left(x_{\uparrow}N\right)=0\quad\mbox{and}\quad\lambda^{-}\left(x_{\downarrow}N\right)=0.
\end{equation}
These points were highlighted in Fig.~\ref{fig:pa-regimes} (d).
We once again treat the special brackets $\left[z\right]_{+}$ as
if they were ordinary and obtain the following non--trivial solutions
for the respective equations:
\begin{align}
x_{\uparrow} & =\frac{q-N^{\beta-1}\sigma_{1}}{q+N^{\beta-\alpha}h},\quad x_{\downarrow}=\frac{h+N^{\alpha-1}\sigma_{0}}{h+N^{\beta-\alpha}q}.
\end{align}
Note that these points are also unstable fixed points.

It is worth to note that not all critical points and regimes are visible
in Fig.~\ref{fig:pa-phases}. In all three cases for $\sigma_{0}=\sigma_{1}$
we have $q_{c,1}=q_{c,2}$ and we show just $q_{c,2}$ in the figure.
For $\alpha=0$ and $\beta=1$ in the infinite $N$ limit $q_{c,3}$
goes to infinity. Even for finitely large $N$ it would be hard to
justify values of $q$ of the order of $q_{c,3}$. Similarly for $\beta=0$
both $q_{c,1}$ and $q_{c,2}$ go to zero in the infinite $N$ limit
and are not shown in Fig.~\ref{fig:pa-phases}.

Qualitatively bifurcation diagrams of the model driven by (\ref{eq:pa-rates})
are quite similar to each other. It appears that both the hybrid case
and the non--extensive case have bifurcation diagrams, which are
essentially appropriately stretched bifurcation diagram for the case
$\alpha=\beta=1$. There is just a single major exception for $q<q_{c,1}$
in $\alpha=0$ and $\beta=1$ case. In this case the model converges
not to a fixed point, but to a stationary distribution. Note that
this regime would be alsp observed in $\alpha=\beta=0$ case, but
in this case $q_{c,1}\rightarrow0$ as $N\rightarrow\infty$.

\section{Support deterring imitation\label{sec:model2}}

Now let us assume that support provided by the allies discourages
only imitation. This means that independent behavior is not affected
by the support and the transition rates take the following form:
\begin{align}
\lambda_{p}^{+} & =\left(N-X\right)\left(\sigma_{1}+\left[\frac{h}{N^{\alpha}}X-\frac{q}{N^{\beta}}\left(N-X\right)\right]_{+}\right),\nonumber \\
\lambda_{p}^{-} & =X\left(\sigma_{0}+\left[\frac{h}{N^{\alpha}}\left(N-X\right)-\frac{q}{N^{\beta}}X\right]_{+}\right).\label{eq:pa2-rates}
\end{align}

\begin{figure}
\begin{centering}
\includegraphics[width=0.4\textwidth]{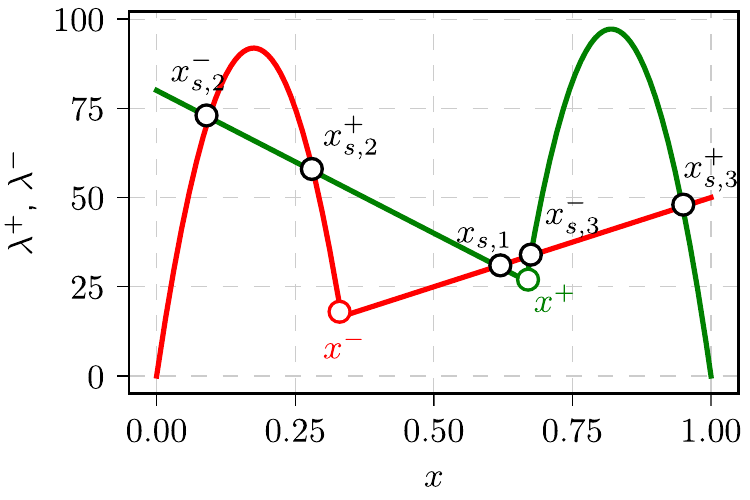}
\par\end{centering}
\caption{Roots of the special brackets (colored circles) and fixed points (black
circles) of the model driven by Eq.~(\ref{eq:pa2-rates}). Green
curve correspond to the birth rate (green curve), while red the death
(red curve) rates. Assumed model parameters: $N=1000$, $\alpha=\beta=1$,
$h=1$, $\sigma_{0}=0.05$, $\sigma_{1}=0.08$.}

\label{fig:pa2-rates}
\end{figure}

To supplement further discussion on the fixed points of this model
let us find the zeros of the special brackets:
\begin{align}
\frac{h}{N^{\alpha}}Nx^{+}-\frac{q}{N^{\beta}}N\left(1-x^{+}\right)=0,\quad & \Rightarrow\quad x^{+}=\frac{N^{\alpha}q}{N^{\beta}h+N^{\alpha}q},\\
\frac{h}{N^{\alpha}}N\left(1-x^{-}\right)-\frac{q}{N^{\beta}}Nx^{-}=0,\quad & \Rightarrow\quad x^{-}=\frac{N^{\beta}h}{N^{\beta}h+N^{\alpha}q}.
\end{align}
We can introduce the four types of fixed points in relation to these
roots. An example with three types of fixed points is provided in
Fig.~\ref{fig:pa2-rates}.

Fixed point of the first type is found when both of the special brackets
evaluate to zero, namely if $x^{-}\leq x_{s,1}\leq x^{+}$. As the
special brackets are zero, the fixed point is obtained by solving:
\begin{equation}
\left(1-x_{s,1}\right)\sigma_{1}=x_{s,1}\sigma_{0}\quad\Rightarrow\quad x_{s,1}=\frac{\sigma_{1}}{\sigma_{0}+\sigma_{1}}.
\end{equation}
This fixed point is stable whenever it is observed. Fixed point of
the first type and fixed point of the fourth type are mutually exclusive.

Fixed points of the second and third types are found when one of the
special brackets evaluates to zero. Therefore we have the following
conditions $x_{s,2}^{\pm}<x^{+}\land x_{s,2}^{\pm}<x^{-}$ and $x_{s,3}^{\pm}>x^{+}\land x_{s,3}^{\pm}>x^{-}$
to observe the respective fixed points. The equations to find fixed
points of the second and third types are given by:
\begin{align}
N\left(1-x\right)x_{s,2} & =\lambda_{p}^{-}\left(x_{s,2}N\right),\\
\lambda_{p}^{+}\left(x_{s,3}N\right) & =Nx_{s,3}\sigma_{0}.
\end{align}
We treat the remaining special brackets as if they were ordinary brackets
and obtain the following solutions to these equations:
\begin{align}
x_{s,2}^{\pm} & =\frac{A\pm\sqrt{A^{2}-2B\sigma_{1}}}{B},\\
x_{s,3}^{\pm} & =\frac{A+2\left(N^{1-\beta}q-\sigma_{0}-\sigma_{1}\right)\pm\sqrt{A^{2}-2B\sigma_{0}}}{B}.
\end{align}
Here we have introduced shorthands for expressions which repeat in
the obtained solutions: 
\begin{equation}
A=N^{1-\alpha}h+\sigma_{0}+\sigma_{1},\quad B=2N\left(N^{-\alpha}h+N^{-\beta}q\right).
\end{equation}
For the fixed points of the second type smaller solution is stable
and the larger one is unstable, while for the fixed points of the
third type larger solution is stable and the smaller one is unstable.

\begin{figure}
\begin{centering}
\includegraphics[width=0.9\textwidth]{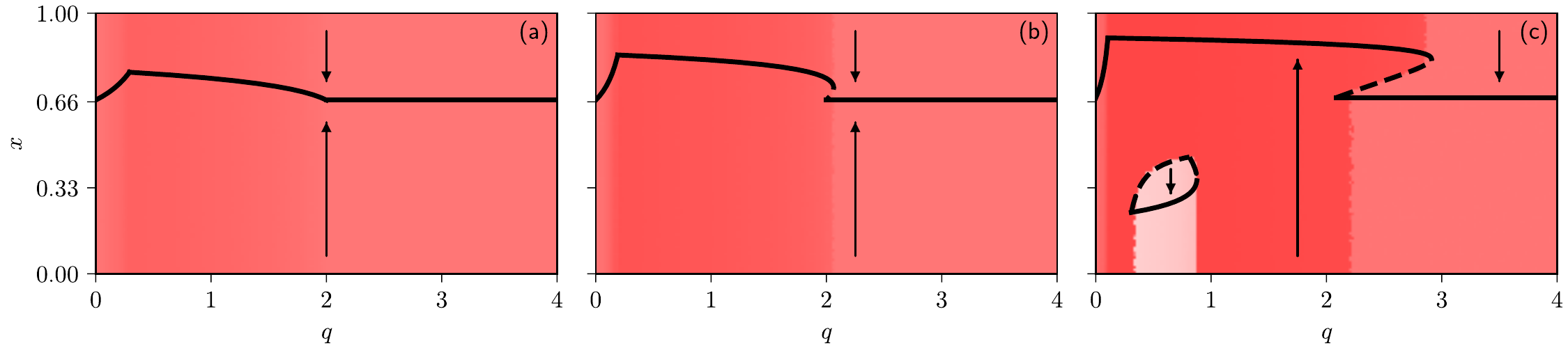}
\par\end{centering}
\caption{Bifurcation diagrams of the model driven by Eq.~(\ref{eq:pa2-rates})
in respect to the intensity of support $q$ in the extensive ($\alpha=\beta=1$)
interaction case. Black curves correspond to stable fixed points (solid
curves) and unstable fixed points (dashed curves). Arrows indicate
directions in which the model moves. Areas are shaded according to
the value of $x$ reached after sufficiently long time (white corresponds
to $x\left(t\right)=0$, red corresponds to $x\left(t\right)=1$,
while shades correspond to some intermediate values). Model parameters:
$N=10^{3}$, $h=1$ (all cases), $\sigma_{0}=0.5$ and $\sigma_{1}=1$
(a), $\sigma_{0}=0.25$ and $\sigma_{1}=0.5$ (b), $\sigma_{0}=0.12$
and $\sigma_{1}=0.25$ (c).}

\label{fig:pa2-evm}
\end{figure}

\begin{figure}
\begin{centering}
\includegraphics[width=0.9\textwidth]{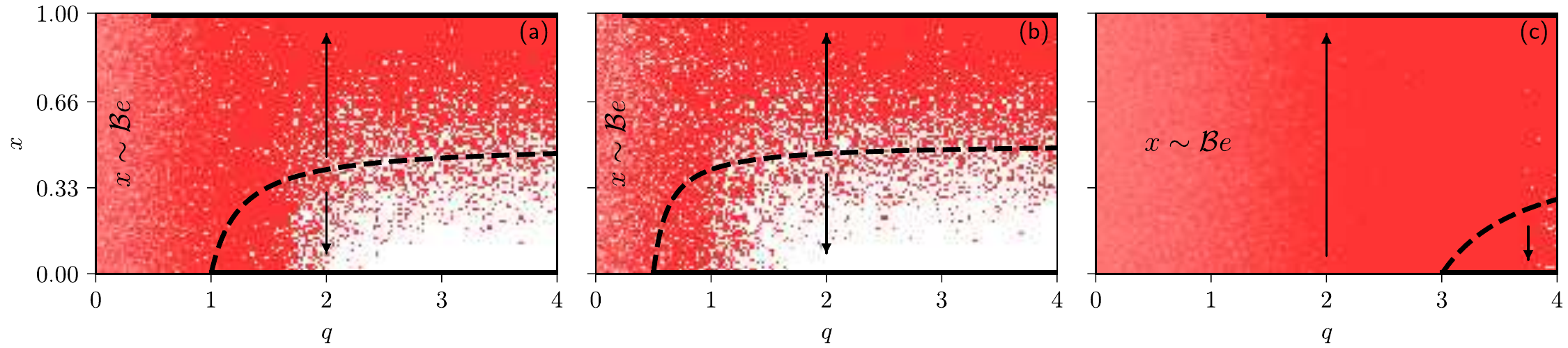}
\par\end{centering}
\caption{Bifurcation diagrams of the model driven by Eq.~(\ref{eq:pa2-rates})
in respect to the intensity of support $q$ in the hybrid ($\alpha=0$
and $\beta=1$) interaction case. Black curves correspond to stable
fixed points (solid curves) and unstable fixed points (dashed curves).
Arrows indicate directions in which the model moves. Areas are shaded
according to the value of $x$ reached after sufficiently long time
(white corresponds to $x\left(t\right)=0$, red corresponds to $x\left(t\right)=1$,
while shades correspond to some intermediate values). Model parameters:
$N=10^{3}$, $h=1$ (all cases), $\sigma_{0}=0.5$ and $\sigma_{1}=1$
(a), $\sigma_{0}=0.5$ and $\sigma_{1}=0.25$ (b), $\sigma_{0}=1.5$
and $\sigma_{1}=3$ (c).}

\label{fig:pa2-hvm}
\end{figure}

\begin{figure}
\begin{centering}
\includegraphics[width=0.9\textwidth]{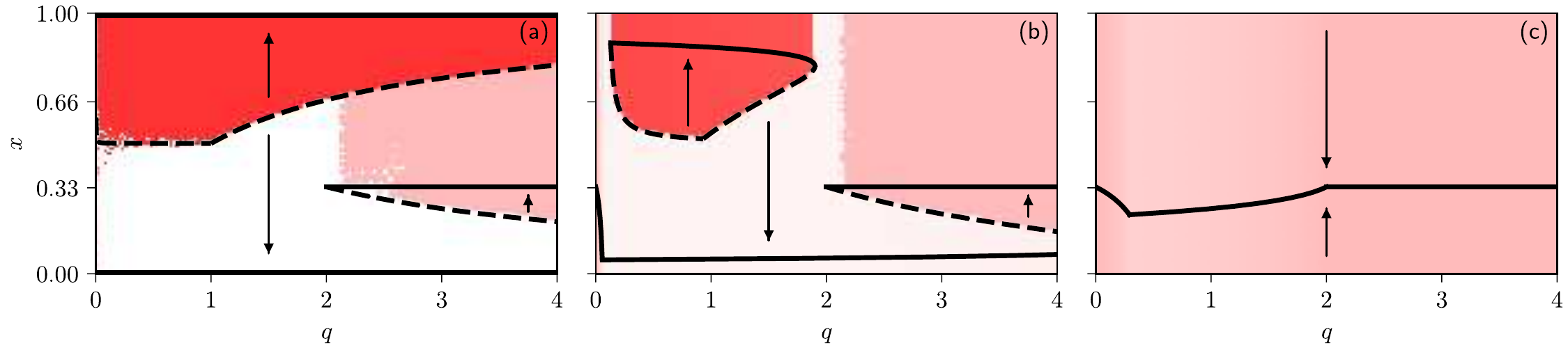}
\par\end{centering}
\caption{Bifurcation diagrams of the model driven by Eq.~(\ref{eq:pa2-rates})
in respect to the intensity of support $q$ in the non--extensive
($\alpha=\beta=0$) interaction case. Black curves correspond to stable
fixed points (solid curves) and unstable fixed points (dashed curves).
Arrows indicate directions in which the model moves. Areas are shaded
according to the value of $x$ reached after sufficiently long time
(white corresponds to $x\left(t\right)=0$, red corresponds to $x\left(t\right)=1$,
while shades correspond to some intermediate values). Model parameters:
$N=10^{3}$, $h=1$ (all cases), $\sigma_{0}=1$ and $\sigma_{1}=0.5$
(a), $\sigma_{0}=120$ and $\sigma_{1}=60$ (b), $\sigma_{0}=10^{3}$
and $\sigma_{1}=500$ (c).}

\label{fig:pa2-nvm}
\end{figure}

Fixed point of the fourth type is found when both of the special brackets
are positive, namely if $x^{+}\leq x_{s,4}\leq x^{-}$. The respective
equation is given by:
\begin{equation}
\lambda_{p}^{+}\left(x_{s,4}N\right)=\lambda_{p}^{-}\left(x_{s,4}N\right).
\end{equation}
Here we treat the special brackets as if they were ordinary ones and
obtain:
\begin{equation}
x_{s,4}=\frac{Nq-N^{\beta}\sigma_{1}}{2Nq-N^{\beta}\left(\sigma_{0}+\sigma_{1}\right)}.
\end{equation}
This fixed point might be stable or unstable depending on the shapes
of the transition rates. Besides the discussed conditions all of the
fixed points must be real and within $\left[0,1\right]$ interval.
Following this discussion one could provide closed form expressions
for the critical points as well as conditions for their existence.
Yet those expressions are rather lengthy and complicated, while the
intuition they provide is very simple and can be shown visually (see
Figs.~\ref{fig:pa2-evm}, \ref{fig:pa2-hvm} and \ref{fig:pa2-nvm}).

As in the previous section we observe that the bifurcation diagrams
of the model driven by Eq.~(\ref{eq:pa2-rates}) are qualitatively
the same for all three considered cases. The bifurcation diagrams
appear to be just the rescaled versions of each other. For large $\sigma_{i}$
only one stable fixed point is observed (as can be seen in Fig.~\ref{fig:pa2-evm}~(a)
or Fig.~\ref{fig:pa2-nvm}~(c)). As $\sigma_{i}$ becomes smaller
an unstable fixed point emerges to act as a separator between two
stable fixed points (see Fig.~\ref{fig:pa2-evm}~(b)). As $\sigma_{i}$
becomes even smaller another pair of stable and unstable fixed points
emerges (see Fig.~\ref{fig:pa2-evm}~(c)). Further decreasing $\sigma_{i}$
triggers emergence of a regime with five fixed points (see Fig.~\ref{fig:pa2-nvm}~(b)
and (a)).

For the hybrid interaction case, Fig.~\ref{fig:pa2-hvm}, we observe
regime, for $q<\min\left(\sigma_{0},\sigma_{1}\right)$, in which
the model is converges not to a fixed point, but to a stationary distribution.
The same regime is observed for the non--extensive interaction case,
too, but in the infinite $N$ limit this regime disappears completely
as $q<\min\left(\sigma_{0},\sigma_{1}\right)/N$ must hold for the
regime to be observed. In both cases the stationary distribution is
the Beta distribution (Beta--binomial for finite $N$) with the following
parameters:
\begin{equation}
x\sim\mathcal{B}e\left(\frac{\sigma_{1}-qN^{1-\beta}}{h+qN^{-\beta}},\frac{\sigma_{0}-qN^{1-\beta}}{h+qN^{-\beta}}\right).
\end{equation}
The stationary distribution is exactly the same as for the model from
the previous section.

For the hybrid interaction case, Fig.~\ref{fig:pa2-hvm}, we also
observe some deviations from the expected behavior. These deviations
are caused by a strong diffusion, which is able to overcome influence
of drift for the smaller $q$ values. Strong diffusion allows trajectories
to escape smaller basins of attraction. Usually those smaller basins
attract to the state having smaller $\sigma_{i}$.

As previously we observe special transitional regime for the extensive
interaction case with $\sigma_{0}=\sigma_{1}=q$. But for this model
the drift function goes to zero for $x^{+}\leq x\leq x^{-}$, while
the trajectories starting outside this interval are attracted to $x^{+}$
or $x^{-}$. For finite $N$ trajectories meander inside the interval,
but in the infinite $N$ limit all the points inside the interval
become stable.

\section{Conclusions\label{sec:Conclusions}}

We have conducted a thorough analysis of the noisy voter model with
supportive interactions. We have considered two different ways to
implement supportive interactions. In the first considered model the
support discourages both independent and peer pressure induced transitions.
In the second model we have assumed that the support discourages only
peer pressure induced transitions. In the first model supportive independent
agents (society of educated people) effectively behave as ordinary
imitative agents (society of uneducated people). Namely the first
model is often driven to full consensus state. Unless support is extremely
strong, in that case polarized states become stable. In contrast the
second model allows partial consensus states to be stable fixed points.
For certain parameter sets there are multiple stable fixed points
corresponding to partial consensus states in the second model.

We have considered both of these models in different agent interaction
scenarios: extensive, non--extensive and hybrid. The extensive and
the non--extensive interaction scenarios were already explored in
\cite{Alfarano2005CompEco,Alfarano2008Dyncon,Kononovicius2014EPJB}.
It was shown that the extensive interactions lead to a macroscopic
description by an ODE (thus the extensive noisy voter model converges
to a fixed point) and the non--extensive interactions lead to a macroscopic
description by an SDE (thus the non--extensive noisy voter model
converges to a stationary distribution). Introducing supportive interactions
into the non--extensive noisy voter model increases impact of drift
and makes diffusion negligible by comparison. Therefore the modified
non--extensive model is approximated by an ODE and converges to a
fixed point. Stationary distribution is observed only in the hybrid
interaction scenario (interactions with the opposition are non--extensive,
while interactions with allies are extensive) when support is weak.

There are other ways to model support in context of the noisy voter
model. One of the possibilities is to have supportiveness depend on
the current system state (i.e., becoming stronger if a group is in
minority). Impact of support could be studied in the $q$-voter model,
which is well--known generalization of the voter model \cite{Castellano2009RevModPhys,Jedrzejewski2019CRP}.
Similarity with non--Markovian opinion freezing models \cite{Stark2008PRL,Stark2008ACS,Wang2014SciRep,Artime2018PRE}
prompts a question to which extent the supportive interactions with
past selves (memory) is equivalent to the supportive interactions
among peers, providing grounds for further inquiry into the nature
of spurious long--range memory \cite{Lanouar2011IJBSS,Gontis2017Entropy}.
It might be reasonable to model support using the complex contagion
framework \cite{Baronchelli2018RSOS}, which would provide a more
transparent view of pairwise interactions between agents in real time.

\section*{Acknowledgements}

Research was funded by European Social Fund (Project No 09.3.3-LMT-K-712-02-0026).


\end{document}